\documentclass[%
 reprint,
 amsmath,amssymb,
 aps,
]{revtex4-2}

\usepackage{graphicx}
\usepackage{dcolumn}
\usepackage{bm}

\begin{document}

\preprint{APS/123-QED}

\title{Brittle fracture studied by ultra-high speed synchrotron X-ray diffraction imaging }

\author{Antoine Petit}
\email{antoine.petit@cea.fr}
\affiliation{Univ. Grenoble Alpes, CEA, LETI, MINATEC Campus, F-38054 Grenoble, France}

 
\author{Sylvia Pokam}%
\affiliation{Univ. Grenoble Alpes, CEA, LETI, MINATEC Campus, F-38054 Grenoble, France}


\author{Frédéric Mazen}
\affiliation{Univ. Grenoble Alpes, CEA, LETI, MINATEC Campus, F-38054 Grenoble, France}
 
\author{Samuel Tardif}
\affiliation{Univ. Grenoble Alpes, CEA, IRIG-MEM-NRS, F-38000 Grenoble, France}

\author{Didier Landru}
\affiliation{SOITEC, Parc Technologique des Fontaines, 38190 Bernin, France}
\author{Oleg Kononchuk}
\affiliation{SOITEC, Parc Technologique des Fontaines, 38190 Bernin, France}

\author{Nadia Ben Mohamed}
\affiliation{SOITEC, Parc Technologique des Fontaines, 38190 Bernin, France}
\author{Margie P. Olbinado}
\affiliation{ESRF – The European Synchrotron, 71 avenue des Martyrs, F-38043 Grenoble, France}

\author{Alexander Rack}
\affiliation{ESRF – The European Synchrotron, 71 avenue des Martyrs, F-38043 Grenoble, France}

\author{François Rieutord}
\affiliation{Univ. Grenoble Alpes, CEA, IRIG-MEM-NRS, F-38000 Grenoble, France}

\date{\today}

\begin{abstract}
Ever since the very first human-made knapped tools, the control of fracture propagation in brittle materials has been a vector of technological development. Nowadays, a broad range of applications relies on crack propagation control, from the mitigation of damages, e.g., from impacts in glass screens or windshields, to industrial processes harnessing fracture to achieve clean cuts over large distances. Yet, studying the fracture in real time is a challenging task, since cracks can propagate up to a few km/s in materials that are often opaque. 
Here, we report on the \textit{in situ} investigation of cracks propagating at up to 2.5 km/s along a (001) plane of a silicon single crystal, using X-ray diffraction megahertz imaging with intense and time-structured synchrotron radiation. 
The studied system is based on the Smart Cut\texttrademark~process, where a buried layer in a material (typically Si) is weakened by micro-cracks and then used to drive a macroscopic crack ($10^{-1}$m) in a plane parallel to the surface with minimal deviation ($10^{-9}$m). 
The results we report here provide the first direct confirmation that the shape of the crack front is not affected by the distribution of the micro-cracks, which had been a postulate for previous studies based on post-fracture results. We further measured instantaneous crack velocities over the centimeter-wide field-of-view, which had only been previously inferred from sparse point measurements, and evidence the effect of local heating by the X-ray beam. Finally, we also observed the post-crack movements of the separated wafer parts, which can be explained using pneumatics and elasticity. Thus, this study provides a comprehensive view of controlled fracture propagation in a crystalline material, paving the way for the \textit{in situ} measurement of ultra-fast strain field propagation.
\end{abstract}

\maketitle

\section{\label{intro}Introduction}
Fracture dynamics and crack propagation in brittle materials are of paramount importance in fields ranging from geology to civil engineering or technological materials science. Frequently, experimental \textit{in situ} studies represent a challenge, in particular in opaque materials, and one is often limited to a \textit{post mortem} analysis \cite{ravi-chandar_dynamic_1998}. Indeed, the spatial extension of the crack in a brittle material is small, with very rapidly varying strain and stress fields near the crack position. In addition, the crack propagation velocity is very high, typically of the order of the Rayleigh velocity or the sound velocity in the material (several km/s). Hence, obtaining a submillimeter resolution without motion blur requires exposure times much below the microsecond. These stringent requirements can only be met using bright synchrotron radiation: the pulsed source enables snapshot images with nanosecond exposure time and hence, freeze ultra-fast motion in the images. 

X-ray scattering, and in particular X-ray diffraction, is a powerful technique whenever crystalline materials are to be studied. The materials penetration properties of X-rays are also well suited to investigate phenomena occurring in the bulk. Diffraction-based imaging, i.e., X-ray (Bragg) topography, takes advantage of the contrast stemming from defects in single crystals to image them in real space, yielding quantitative information, as recently reviewed by \cite{danilewsky_x-ray_2020}. In the field of fracture mechanics, X-ray topography has proven to be an efficient tool to investigate cracks in single crystal \cite{tanner_prediction_2012, tanner_x-ray_2013, tanner_geometry_2015, danilewsky_crack_2013, atrash_crystalline_2017}. Numerous studies have been conducted to measure the strain and estimate the stress near crack tips, but they usually only dealt with static or quasi-static distributions of strain and stress, as discussed in an extensive review \cite{withers_fracture_2015}. Combined with hard synchrotron radiation, X-ray topography can be performed in real-time \cite{tuomi_real-time_1983, rack_possibilities_2010}, under mechanical \cite{tanner_quantitative_2017, tsoutsouva_dynamic_2021} or thermal loading \cite{danilewsky_real-time_2011, danilewsky_crack_2013}. More recently, some of the present authors have shown that thermally induced crack propagation could be followed \textit{in situ} using X-ray imaging, both in transmission and diffraction conditions, with peak velocities observed on the order of a few m/s \cite{rack_real-time_2016}. We show in this report that synchrotron X-ray diffraction imaging can actually be fast enough for the real time and real space study of fracture mechanics near the sound velocity, providing unprecedented direct observations of crack propagation and a wealth of information at different time and length scales, otherwise inaccessible.

This report focuses on single crystal silicon, the most common substrate for the microelectronics industry. In particular, our test vehicles are assemblies of silicon wafers prepared with the Smart Cut\texttrademark~technology. This process is widely used to transfer thin single crystal layers to a different substrate by fracturing in a controlled manner an implanted interface \cite{bruel_silicon_1995}. The transfer can be done on any substrate and this technology is currently used for the mass production of Silicon On Insulator (SOI) substrates \cite{di_cioccio_silicon_1997, tauzin_transfers_2005}. The full workflow of the technology is described elsewhere and can be summarized as follows. First, a high dose of light ions, typically hydrogen \cite{bruel_silicon_1995, aspar_basic_1997}, or a mixture of hydrogen and helium \cite{agarwal_efficient_1998, duo_defect_2000}, is implanted in a silicon substrate. This results in the formation of a buried weakened layer within the substrate. By bonding a stiff substrate on the surface of the implanted wafer (e.g., an “acceptor” wafer), the implantation-related defects grow in-plane under annealing, forming a network of pressurized microcracks \cite{moriceau_smart_2012, claverie_thermal_2018}. The size of these microcracks depends on the annealing conditions and the nature and amount of implanted gases. A typical image obtained by infrared (IR) confocal microscopy is shown in Figure \ref{fig1}b, showing an assembly of flat microcracks with typical sizes in the few µm range after few hours of annealing at 375°C. Once the surface coverage of the microcracks is large enough, a manually generated or naturally occurring fracture propagates through the weakened layer and effectively transfers the surface film to the acceptor \cite{bruel_silicon_1995}. 

Before the fracture, the system is in a metastable state where the interface is internally loaded by the distribution of pressurized microcracks. Each microcrack is in pseudo-equilibrium, as its evolution is very slow and the equilibrium state can be described using the standard Griffith argument, where the energy release rate is equal to the surface energy \cite{penot_development_2013}. The internal load on the microcrack faces is due to the inner pressurized gas, which has precipitated from the implanted material layer. The pressure load is compensated by the elastic deformation of the material and the surface opening. The implanted dose is usually around $10^{16}$~$at/cm^{-2}$, therefore the pressure in a micrometer-sized crack is on the order of 100 MPa, and the corresponding microcrack height is in the nanometer range. A crude description of each microcrack is the so-called penny-shape crack model that can be found in standard textbooks \cite{tada_stress_1985}. 

Upon fracture, the crack front propagates in the microcracks layer, akin to a 2-dimensional pre-perforated material. The crack propagation plane is close to the middle of the 2 × 775 µm thick silicon wafer assembly, parallel to the wafer external surfaces, i.e., in the so-called strip geometry. It has been shown that in this case, the crack velocity reaches rapidly an asymptotic velocity close to the Rayleigh velocity \cite{massy_fracture_2015} which prevents a deflection of the front \cite{kermode_low-speed_2008}. The crack front emits acoustic waves and the feedback of those acoustic waves on the propagating crack front causes small deviations of the fracture plane \cite{marder_new_1991, landru_fracture_2021}. This translates into periodic patterns of roughness modulations on the surfaces obtained after splitting \cite{massy_fracture_2015, massy_crack_2018}. Similar results have also been found in asperity-free silicon samples \cite{wang_self-emitted_2020}. Therefore, the study of the propagation is key to understand the dynamic brittle fracture in such structures and its impact on the morphology of the separated surfaces.

As mentioned earlier, the \textit{in situ} experimental study of crack propagation is difficult in opaque, brittle materials. Infrared lasers have been used to measure the crack velocity \cite{massy_fracture_2015} and crack opening displacement \cite{massy_crack_2018}, but this technique only gives a point measurement at the laser spot position. To the best of our knowledge, no direct imaging of the crack front propagation at Rayleigh velocity in silicon substrates has been reported so far. Here, we fill this gap by presenting an original approach taking advantage of short X-ray synchrotron pulses and an ultra-high-speed X-ray detector to illuminate the Si crystal in diffraction conditions and consequently, image the crack at different steps of its propagation. Our direct observation of crack propagation provides measurements of both the crack front shape and velocity, together with post-split wafer deformation. 

\section{\label{methods}Methods}

\subsection{\label{exp setup}Experimental setup}
\begin{figure*}
\includegraphics[width=\textwidth]{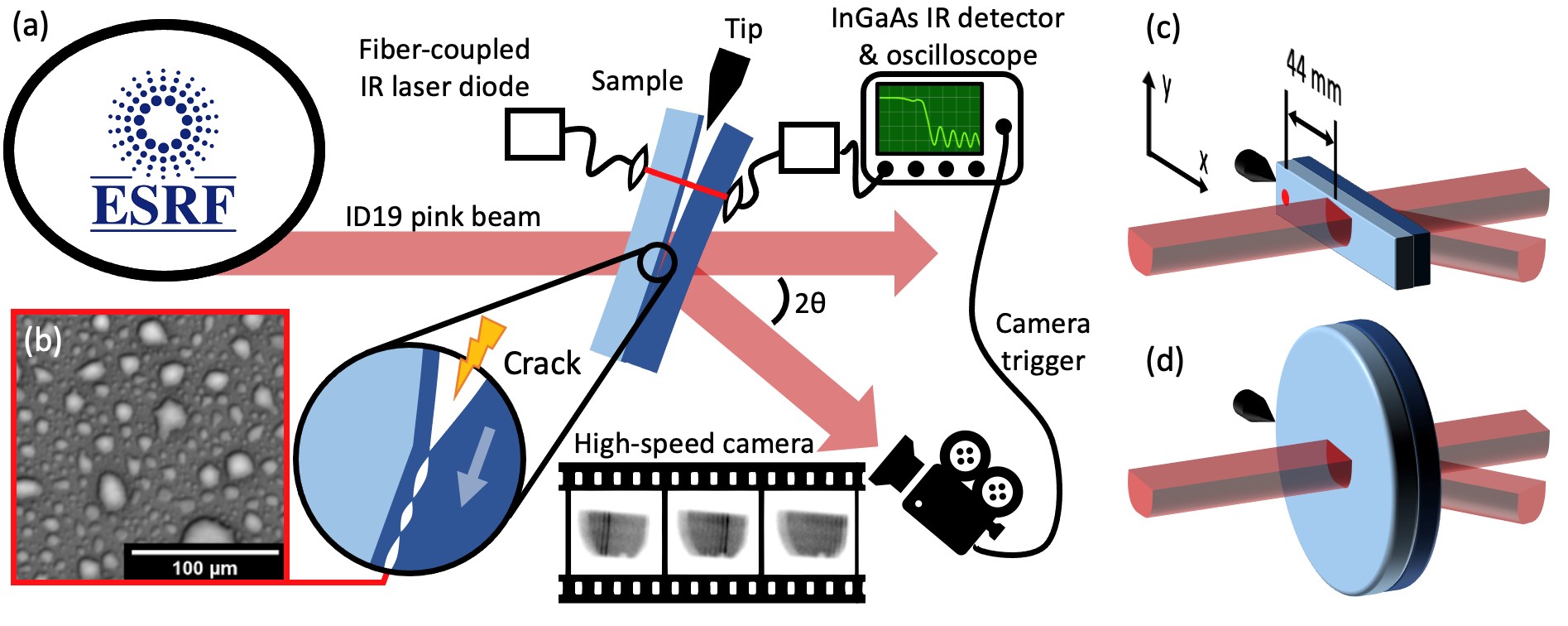}
\caption{(a) Top view of the imaging experiment setup used at ID19 beamline, the ID19 pink beam illuminate the sample (bonded wafers) in diffraction condition, and high-speed camera connected to an image intensifier records the diffracted image on the scintillator. The motorized tip is used to mechanically initiate the fracture, which is detected by the infra-red laser, which in turn is used to trigger the camera. The zoom at the center of the sample shows the fracture propagation through the pressurized micro-cracks layer. (b): Top view of the pressurized micro-cracks layer, made by implantation-related defects after a few hours of annealing. (c, d): 3D representation of the strip and wafer samples, respectively. The red dot represents the IR laser.}
\label{fig1}
\end{figure*}

For this experiment, ultra-high speed diffraction imaging is deployed at the ID19 beamline at the ESRF, as shown in Figure \ref{fig1}a \cite{weitkamp_status_2010, escauriza_ultra-high-speed_2018}. The X-ray setup details of the ID19 beamline follow a protocol for high-speed diffraction imaging of wafer fracture \cite{rack_real-time_2016}. The beamline undulators were optimized for 20 keV photon energy, the impinging radiation on the wafer was basically white with only a diamond and an aluminum attenuator in the beam. We could illuminate a wide area of the sample with a large and parallel size beam thanks to the long distance between the source and sample (150 m). The typical beam size at the sample position was a rectangle of 21 mm * 12.5 mm. For the experiment, the ESRF operated in the so-called timing mode 4-bunch where 4 highly populated electron bunches (10 mA) are stored equidistantly in the storage ring (100 ps pulse duration, 704 ns pulse period \cite{rutherford_evaluating_2016}). 

Two types of samples were used. The starting material was assemblies of two [001]-oriented 300 mm diameter, 775 µm thickness silicon wafers, prepared by implantation of hydrogen and helium at doses around $10^{16}$~$at/cm^{-2}$, direct bonding and annealing to form a buried layer of micro-cracks (Figure \ref{fig1}b). They were then either cut along the radius in 20 mm-wide strips, with their length along [110] or [100] (Figure \ref{fig1}c), or used as they were (Figure \ref{fig1}d). 

The samples were mounted vertically on a rotation stage such that their [110] or [100] direction (i.e. their length for the strip samples) was horizontal. They were then rotated around a vertical axis to bring the (220) or (400) planes in diffraction condition in the horizontal scattering plane, in the so-called Laue transmission geometry. For the full wafer sample, the 220 diffraction was chosen. 

Initially, the intensity received on the detector is constant since the two wafers are still bonded and do not move. When the crack front crosses the beam-illuminated area, the transmitted intensity changes due to the assembly opening following the crack front, thus imaging the crack front. The scattered X-ray beam image is converted into visible light using a combination of a scintillator and an X-ray image intensifier \cite{ponchut_evaluation_2001}. The image was recorded using an ultrafast camera type HPV-X2 (Shimadzu Corporation, Japan) \cite{olbinado_mhz_2017}, placed 170 cm downstream from the sample, and offset by the scattering angle ($\approx$ 18.6° for 220, 26.4° for 400). The camera can acquire 256 frames at a rate up to 10 MHz and exposure time down to 50 ns (depending on the acquisition mode). Here, thanks to the stroboscopic nature of the incident radiation, the effective exposure time is given by the bunch length and not by the integration window of the camera, provided that the integration window is equal or larger than the pulse period (so-called single bunch imaging). We thus used the camera either in 1 MHz mode or 1.4 MHz mode (700 ns and 400 ns integration time, respectively). The effective pixel size of the detector (camera coupled with X-ray image intensifier) is either 300 µm or 165 µm, depending on the magnification ratio selected.

The triggering of the camera was performed using the optical setup described hereafter. An infrared laser beam is directed to an InGaAs IR photodiode through the sample in such a way that the fracture front crosses the laser beam before the X-ray illuminated area. The photodiode signal is recorded by an oscilloscope and a trigger threshold can be set on the transmitted intensity, as the latter changes sharply when the crack front crosses the laser beam. The time delay between the trigger signal and the camera can be adjusted according to the distance between the IR laser spot and the X-ray illuminated region and to the crack velocity. In addition, the crack opening following the crack front creates air wedge interference fringes, which are visible in the transmitted signal from the IR photodiode. These signals can be used for dynamic measurement of the crack velocity and give detailed information on the vibrations of the wafers associated to the crack front propagation, including acoustic (Lamb) waves \cite{massy_fracture_2015, massy_crack_2018}. Finally, in our experiment, the samples were pre-annealed to develop microcracks up to desired microcrack surface coverage and a motor-driven blade was used to mechanically induce the fracture at room temperature. The motorized blade was slowly inserted in the beveled edge of the wafers until the fracture occurs.

\section{\label{results_discussion}Results and discussion}
\subsection{\label{crack_front shape}Crack front shape}
\begin{figure}[b]
\includegraphics[width=\columnwidth]{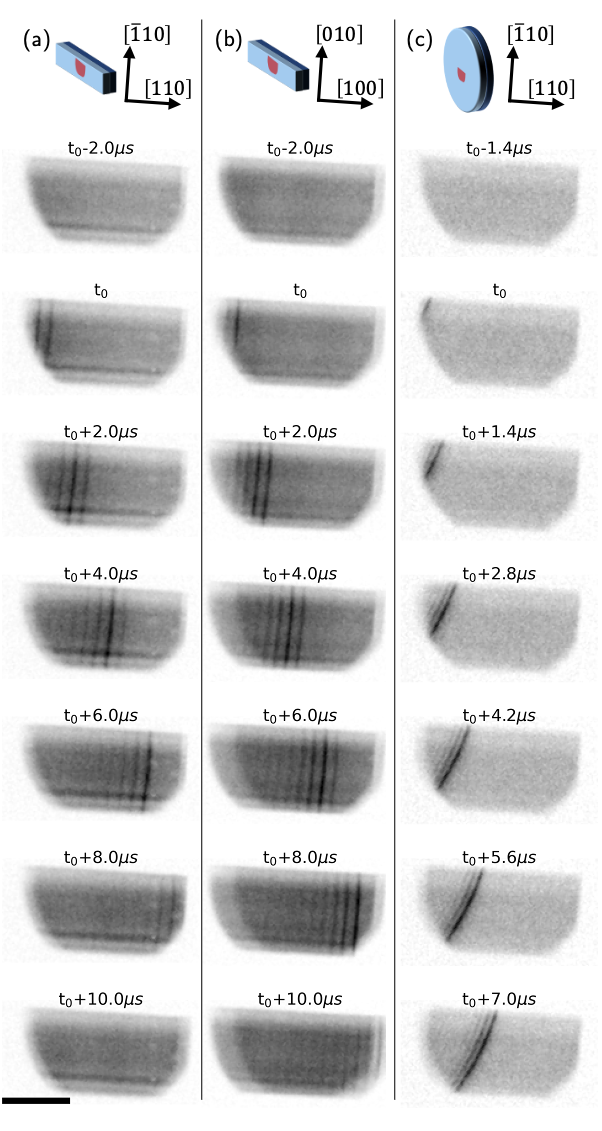}
\caption{X-ray diffraction imaging of the crack front in silicon (a) [110]-strip, (b) [100]-strip and (c) full 300 mm wafers. The time $t_0$ indicates the entry time of the crack front in the field of view. The scale bar is 10 mm.}
\label{fig:fig2}
\end{figure}

The evolution of the diffracted signal (220 or 400 Bragg reflection) recorded by the camera when the crack crossed the field of view is shown in Figure \ref{fig:fig2} for the different types of samples. Figure \ref{fig:fig2}a,b shows the crack propagation in strip samples, while Figure \ref{fig:fig2}c shows the propagation in a full assembly of 300 mm wafers. For the two types of samples, the first image is taken few microseconds before the crack front reaches the illuminated area. Then, the following images are taken at regular interval, in a stroboscopic manner (for clarity, we only show every other image). The comparison of the different delayed images evidences the crack front propagation. As we will show after, the enhanced scattered intensity at the crack front is due to an overlapping of the diffraction of the strongly curved region at the crack front. Thanks to the wide incidence spectrum, the tilted crystal regions around the crack front are still in diffraction condition. Thus, the crack front can be directly visualized as an intense signal line visible at the center of the diffracted area on the images of the Figure \ref{fig:fig2}. The very short illumination due to a single bunch provides a snapshot of the crack front shape with negligible blurring due to crack movement during exposure. For all the samples, the crack front has a smooth circular shape, within the resolution of our setup, which experimentally confirms for the first time what was previously a working hypothesis for the analysis of the crack propagation \cite{landru_fracture_2021}.

\subsection{\label{crack_front velocity}Crack front velocity}
\begin{figure*}
\includegraphics[width=\textwidth]{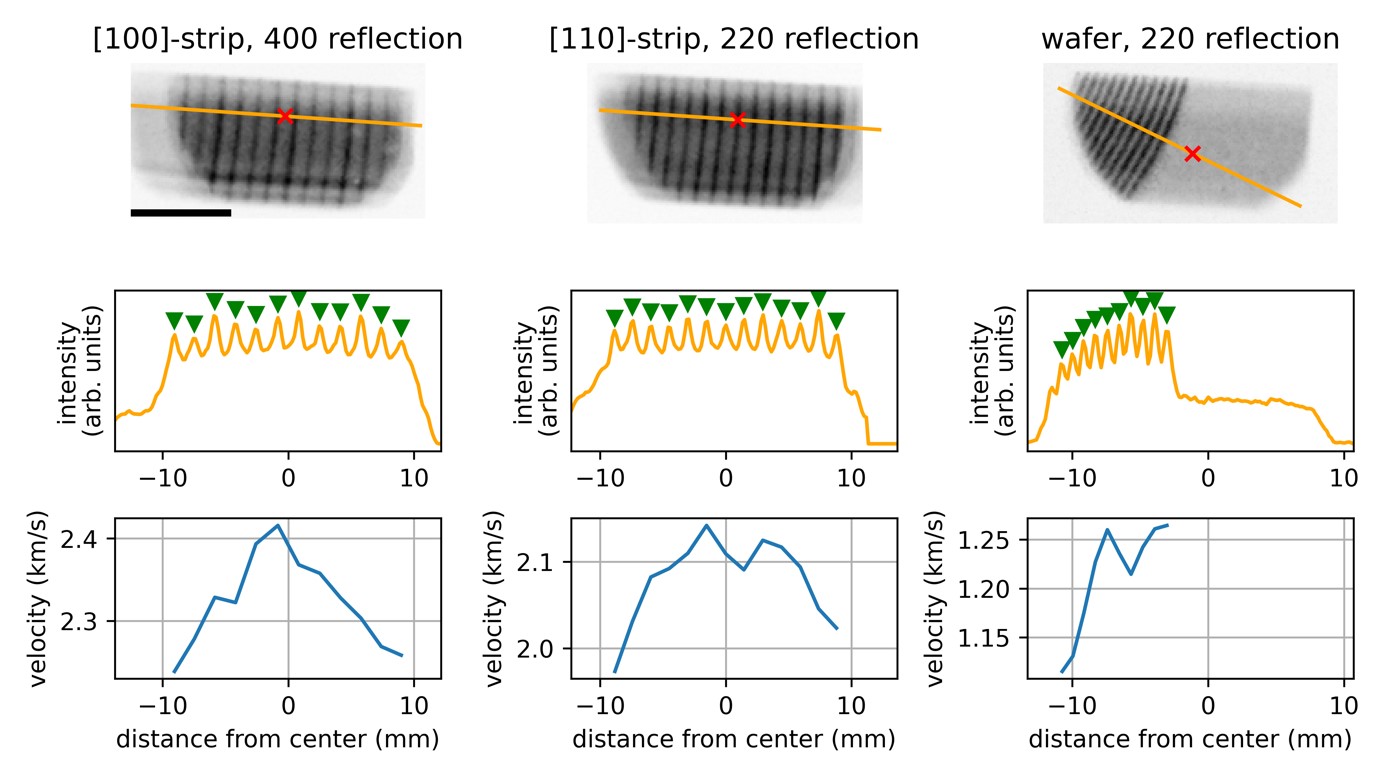}
\caption{(top) single image of all the maximum intensities seen by the camera for the different strip samples and full 300 mm wafers. The center of the X-ray illuminated area is represented by a red cross, the orange line indicates the position of the line profile extracted below. (center) line profile perpendicular to the crack front. Each peak is due to a single bunch. Space-wise, the exact location of each peak is obtained by a gaussian fit and indicated by a green triangle. Time-wise, each bunch is 704 ns after the previous one. (bottom) evolution of crack front velocity in the X-ray illuminated area, as calculated from the peaks space-time positions.}
\label{fig3}
\end{figure*}

Interestingly, the fracture speed does not seem to depend strongly on the crystal orientation, as the topographs from Figure \ref{fig:fig2}a and Figure \ref{fig:fig2}b seem to be quite synchronized. Knowing the size of the illuminated area (21$\pm$ 1 mm) and the travel time for the crack front through this area (9$\pm$ 1 µs) for Figure \ref{fig:fig2}a and Figure \ref{fig:fig2}b, the crack front velocity can be estimated to about 2.4$\pm$0.4 km/s for both strip samples. The part of the signal behind the crack front presents several stripes due to the afterglow of the scintillator associated with the pulsed time structure of the X-ray beam in the 4-bunch mode filling pattern of the ESRF storage ring. Indeed, the scintillator is often responsible of a deleterious background intensity between bunches \cite{rutherford_evaluating_2016}. Here, we harnessed this time structure to obtain a more accurate instantaneous crack velocity. To do so, we compute a max-filtered image for each sample by taking the maximum value observed at any time for every pixel of our set of images. This is somewhat akin to a “bulb exposure” photograph, showing in a single image all the maximum intensities seen by the camera (Figure \ref{fig3}, top row). We then extract a line profile perpendicular to the crack front, showing the peaks due to every single bunch. As mentioned before, the time separation of the bunches is well-known and fixed by the storage ring filling pattern. To quantify their space separation, we fitted each peak locally by a Gaussian function to extract the peak position with sub-pixel resolution (Figure \ref{fig3}, middle row). We could thus compute the instantaneous crack velocity for each bunch (Figure \ref{fig3}, bottom row). 

The crack velocity is slightly higher for the strip sample along [100] than for the strip along [110] (2.2 km/s to 2.4 km/s and 2.0 km/s to 2.15 km/s, respectively). This can be explained in the context of dynamic crack propagation, where the asymptotic crack velocity v is proportional to the Rayleigh wave velocity $v_R$ \cite{freund_dynamic_1990} :
\begin{equation}
v=v_R\left(1-\frac{\Gamma}{G} \right)
\end{equation}
with $\Gamma$ the fracture energy, related to the area between the microcracks where the fracture must break the material, and G the energy release rate, i.e. the internal elastic energy recovered when the material is cracked, here directly related to the pressure inside the microcracks \cite{massy_fracture_2015}. Since the strip sample along [110] and along [100] have been cut out from the same wafer, the microcracks are identical (density, internal pressure) and thus $\Gamma$ and G should identical as well. The Rayleigh velocity depends however on the propagation direction, from 4.90 km/s along [100] to 5.06 km/s along [110] \cite{pratt_acoustic_1969}. Thus, the observation of a faster crack propagation along [100] is not consistent with this description. We also note that the instantaneous velocity for the 300 mm wafer sample is significantly lower. Both of these results are explained hereafter. 

The crack velocity is noticeably higher near the center of the X-ray illuminated area for each strip sample: 2.41 km/s (2.13 km/s) at the center vs 2.24 km/s (1.97 km/s) at the edge of the [100]-strip sample ([110]-strip sample). This can actually be explained by the local heating of the sample due to the intense X-ray beam. As mentioned before, the energy release rate G is proportional to the pressure in the microcracks, and thus to the temperature. The relation $1-v⁄v_f = A/T$, where A is a constant, was already experimentally verified in a previous study \cite{massy_fracture_2015}. Therefore, the observed 8 \% velocity increase at the center of the field of view for both samples corresponds to a 5\% to 6\% temperature increase. This would amount to a $\Delta T=$15 K to 25 K around room temperature (300 K to 400 K). 

Similarly, a different average sample temperature over the field of view can explain the discrepancy in the numerical values between the dynamic model ($v_{110}⁄v_{100} =1.03$) and the experimental value ($v_{110}⁄v_{100} =0.90\pm0.01$). The faster crack velocity along [100] would correspond to a $13\pm1\%$ temperature increase compared to the [110]-strip, i.e. the [100]-strip sample was about 40 K to 50 K hotter. A longer alignment time for the first [100]-strip sample would have resulted in an increased exposure to the beam, and therefore in the increased heating before the fracture was triggered. We note that the samples were too hot to be touched bare handed right after the measurement, indicating temperatures typically above 340 K.

Using the same approach, comparing the 300 mm wafer sample and the strip samples, the respective velocities indicate an average temperature about 30\% higher in the strip samples. This indicates that the strip samples were about 100 K hotter than the wafer sample, which is consistent with their much smaller heat capacity and heat dissipation.

\subsection{\label{post crack oscillations}Post crack oscillations}
As shown in Figure \ref{fig:fig2}, we observe a signal modulation in the wake of the crack front. This is particularly visible on the edges of the beam footprint. The detector image actually shows the superimposition of two scattered images of the incident beam, with a time-dependent shift. These two images are the scattered images by the two arms of the crack opening. The crack opening angular displacement can thus be obtained from the analysis of the diffracted image motions. Given the beam divergence (21 mm over 150m) and bandwidth ($\lambda/\Delta\lambda$=100), we estimate from the Du Mond diagram for the 220 (400) reflection that sample rotations up to about 0.8 (1.1) mrad  can still be imaged (Supplementary Information). This is larger than the typically deflection that we have previously measured using optical reflection \cite{ronseaux_experimental_2021}, indicating that the full movement of the two sample plates after the fracture can be captured.

\begin{figure}
    \centering
    \includegraphics[width=\columnwidth]{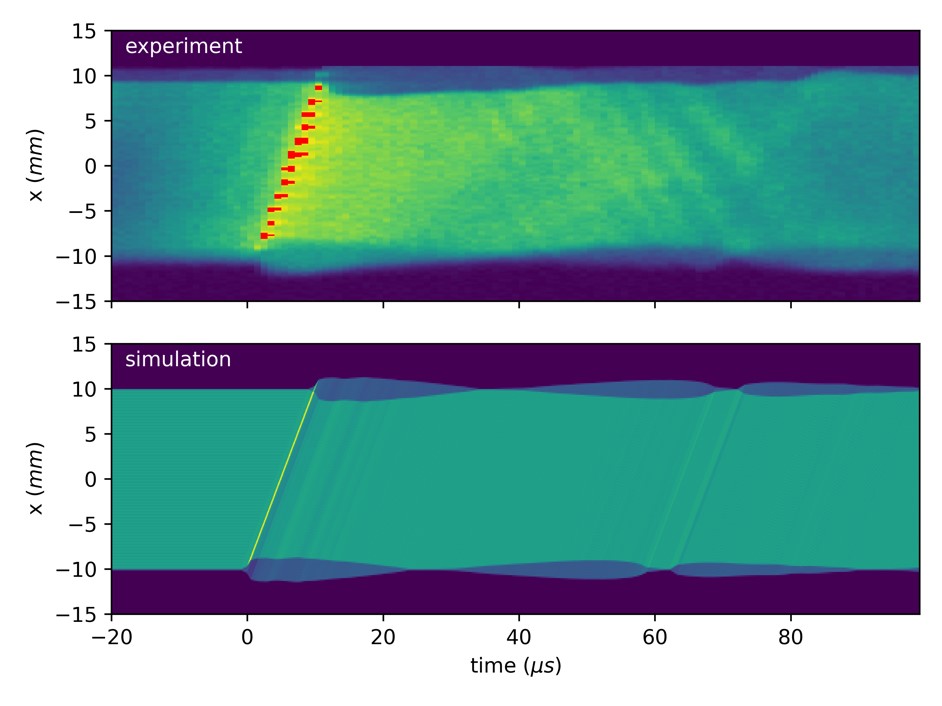}
    \caption{Diffracted intensity for the [110]-strip sample as a function of time and position along the length x (i.e. along [110]), taken at center y position, (top) experimental data, (bottom) simulated data considering the propagation of the gap opening profile. }
    \label{fig:num4}
\end{figure}

To analyze the post-crack sample movement, we extract the center line of the diffracted image (averaged over 5 pixels) and plot it as a function of time in Figure \ref{fig:num4}. The facture propagation from one side to the other is clearly seen as the intense diagonal line with slope $v_f$. The crack opening is directly 
visible in the wake of the crack front. Shortly after the crack front crossing, the two separated plates can be seen as the half-intense areas at each extremity, where only one of the two plates contributes to the diffraction. Note that small similar shadows could be observed on each side before the crack apparition, simply due to a slight misalignment of the crystal planes in the two bonded wafers. The angular separation just behind the crack indicates a “wedge” shape of the crack opening. This is consistent with infrared interferometry measurements of the crack opening \cite{massy_fracture_2015} and optical reflection measurements \cite{ronseaux_experimental_2021}. From linear elastic fracture mechanics, the profile is parabolic in the close vicinity of the crack front, as strain varies as K/$\sqrt{x}$, with K the stress intensity factor. At larger distances, a regular beam bending strain should be expected. The wedge shape is probably due to the transition between these two regimes with opposite curvatures and a convolution by the experimental resolution. The later oscillatory shape finally indicates a back-and-forth movement of the plates.

In order to go further in the analysis, we simulated the diffracted images using the following approach. The IR laser signal to trigger the high-speed camera also provides an indirect measurement of the gap profile behind the crack front. The method is described elsewhere \cite{massy_fracture_2015} and can be summarized as follows. The opening of the crack creates two partially reflecting surfaces. The transmission through this gap is thus a function of the gap opening, with maxima every $\lambda$/2, resulting in fringes in the detected signal as a function of time. It is not directly possible to know if the gap is opening or closing by this quantified amount at each fringe, but it is reasonable to assume that a maximum in the delay between two successive fringes (vanishing speed) indicates a reversal of the motion. We thus obtain the local discrete profile $\zeta(t)$ of the gap at the location of the IR laser, where each plate is deformed by $\pm\zeta(t)$/2 (see Supplementary information). Assuming that the profile is invariant by translation, i.e., the sample deformation behind the crack is the same throughout the crack propagation, $\zeta(t)=\zeta(-x/v_f)$. The angle of the surfaces can then be computed as:
\begin{equation}
    \alpha=\pm\frac{1}{2}\frac{\partial\zeta}{\partial x}=\pm\frac{1}{2v_f}\frac{\partial\zeta}{\partial t}
\end{equation}

The corresponding diffraction image can be simulated using a simple ray-tracing approach. For a time step t, a length of 21 mm of the deformation profile is illuminated by parallel rays every 10 µm. Each ray is then deflected by $\pm2\alpha$, where $\alpha$ is the local sample rotation defined above. The rays propagate over d=1.7 m and hit the detector where all the resulting positions are binned according to the pixel size. The initially illuminated area is shifted for the next time step $t+\Delta t$ by $v_f\Delta t$. The resulting simulated image is shown in Figure \ref{fig:num4} as well. The main features are well captured by the simulation, confirming that our description is correct. The observed back-and-forth motion of the two plates visible on the edges (x=$\pm$10 mm) is actually due to pneumatic oscillations, as described before \cite{massy_fracture_2015}. Small discrepancies can be observed for the later times, which may indicate that the deformation profile is not exactly invariant by translation, probably because of the finite size of the sample and/or local temperature changes. Finally, back propagating waves are observed in the experimental data as slightly more intense lines with a negative slope in Figure \ref{fig:num4}, yet they are not reproduced by our simple simulations. These waves could actually be flexural waves, recently observed using optical reflection measurements \cite{ronseaux_experimental_2021}.

\section{\label{summary_outlook}Summary and outlook}
As it had previously been foreseen \cite{rack_real-time_2016}, X-ray diffraction imaging using a large parallel synchrotron beam as proven to be a unique tool for real-time dynamic fracture studies, thanks to the high-flux and temporal resolution of synchrotron illumination, coupled with high-speed cameras. Here, we showed that synchrotron X-ray diffraction imaging can be fast enough for the in-situ study of fracture mechanics in a brittle material, at speed over few km/s. We harnessed this technique to tackle the industrially-relevant problem of the fracture propagation in the Smart Cut\texttrademark~technology and to obtain otherwise inaccessible data. In particular, both real-space and real-time images of the crack front propagating in single crystal Si at speed near the Rayleigh velocity were acquired for the first time. This direct visualization of the crack front shape experimentally confirmed the homogeneous propagation of the fracture, which was previously only a working hypothesis for the fractography analysis of post-split surfaces. Thanks to the time structure of the synchrotron source, the local crack velocity could be measured and compared to point measurement using infra-red laser transmission. While the average velocity value is consistent with IR measurements, the effect of increased velocity near the image center could be interpreted as a local heating due to the intense X-ray beam, in agreement with previous studies \cite{massy_fracture_2015, landru_fracture_2021}. Finally, the post-split images showed the movements of the separated wafers and confirmed the importance of the pneumatic oscillations in the wake of the fracture. Additional back-propagating waves were also observed, reminiscent of flexural waves and requiring further analysis. 
These results open a vast avenue of future potential studies. For example, it could be very interesting to trigger the fracture thermally using the X-ray heating of the sample, to be as close as possible to the phenomenon occurring in the annealing furnace, and thus observe the very first stages of fracture initiation. A more detailed stress and strain description at the crack tip is now probably at reach, using e.g., a larger sample to detector distance, combined with the improved brilliance of the new ESRF source. Also, coupling high-speed diffraction imaging (few MHz) with more conventional yet fast optical imaging (10 kHz to 100 kHz) could provide a full understanding of the fracture propagation, the related acoustic waves emission and their intimate interplay. Finally, our approach here is not limited to silicon and can be extended to any crystalline material of interest that can be transferred using the Smart Cut\texttrademark~technology, such Ge, GaAs or SiC.

\begin{acknowledgments}
We thank the staff of the ESRF beamlines ID19 and BM32 for dedicated technical assistance setting up and running the experiment. Beam time allocation is acknowledged. The support of the FRAINDY project from the French national research agency under No. ANR-18-CE08-0020 is acknowledged.         
\end{acknowledgments}


%


\end{document}